\documentclass{cmspaper}
\usepackage{color}

\begin{document}
\begin{titlepage}
\date{12 May 2006}


\title{GARCON: Genetic Algorithm for Rectangular Cuts
  OptimizatioN. User's manual for version~2.0.}

  \begin{Authlist}
    S.~Abdullin\Iref{a}, 
    D.~Acosta\Iref{b}, 
    P.~Bartalini\Iref{b},
    R.~Cavanaugh\Iref{b}, 
    A.~Drozdetskiy{\color{red}\Iref{e}}\Iref{b},
    G.~Karapostoli\Iref{c}, 
    G.~Mitselmakher\Iref{b}, 
    Yu.~Pakhotin\Iref{b},
    B.~Scurlock\Iref{b}, 
    M.~Spiropulu\Iref{d}
  \end{Authlist}
  \Instfoot{e}{Corresponding author, e-mail: Alexey.Drozdetskiy@cern.ch}
  \Instfoot{a}{Fermi National Laboratory, Chicago, IL, USA}
  \Instfoot{b}{University of Florida, Gainesville, FL, USA}
  \Instfoot{c}{CERN and University of Athens, Greece}
  \Instfoot{d}{CERN}

  \begin{abstract}
    This paper presents GARCON program, illustrating its functionality
    on a simple HEP analysis example. The program automatically
    performs rectangular cuts optimization and verification for
    stability in a multi-dimensional phase space. The program has been
    successfully used by a number of very different analyses presented
    in the CMS Physics Technical Design Report. The current version
    GARCON~2.0 incorporates the feedback the authors have
    received. User's Manual is included as a part of the note.
  \end{abstract} 

  
\end{titlepage}

\setcounter{page}{2}


\section{Introduction}

Genetic algorithm (GA) definitions along with some review information
are given in Ref.~\cite{GA}. In short, GA is a set of algorithms
inspired by concepts of natural selection with evolving individuals,
which allowed to be created randomly, to mutate, inherit their
qualities, etc. useful in optimization problems with a large number of
discrete solutions.

Typically, a High Energy Physics (HEP) analysis has quite a few selection
criteria (cuts) to optimize for example a significance of the
``signal'' excess over ``background'' events: transverse
energy/momenta cuts, missing transverse energy, angular correlations,
isolation and impact parameters, etc. In such cases simple scan over
multi-dimensional cuts space (especially when done on top of a scan
over theoretical predictions parameters space, e.g. for SUSY) leads to
CPU time demand varying from days to many years... One of the
alternative methods, which solves the issue is to employ a Genetic
Algorithm, see e.g.~\cite{HOLLAND,GOLDBERG,SALAVAT}.

We wrote a code, GARCON~\cite{GARCONMAIN}, which automatically
performs cut optimization and verification for stability
effectively trying $\sim 10^{50}$ cut set parameters/values
permutations for millions of input events in hours time
scale. Examples of analyses with GARCON can be found presented in the
CMS Physics TDR, v.2~\cite{PTDR2} and in recent
papers~\cite{UF_SUSY,UF_h4mu,SINGLETOP}.

In comparison with other automated optimization methods GARCON output
is transparent to user: it just says what rectangular cut values are
optimal and recommended in an analysis. An interpretation of these cut
values is absolutely the same as when one selects a set of rectangular
cut values for each variable in a ``classical'' way ``by eye'', except
in the case of GARCON those cut values would be optimal to deliver the
best value of the function used for optimization\footnote{Hard-coded
popular significance estimators as well as a possibility for a
User defined function, are described in Appen.~\ref{sec:quality}}.

In this paper we describe the basics of the GA, illustrating GARCON
functionality on a simple example of a ``toy'' MC generator-level
analysis. A significant part of the paper consists of User's Manual
describing how to use the program (Sec.~\ref{sec:manual}).

GARCON version 2.0~\cite{GARCONMAIN} among many other features allows
user:

\begin{itemize}

\item to select an optimization function among known significance
  estimators, as well as to define user's own criteria, which may be as
  simple as signal to background ratio, or more complicated, including
  different systematic uncertainties separately on signal
  and background processes, different weights per event, etc.;

\item to define a precision of the optimization;

\item to restrict the optimization using different kind of
  requirements, such us minimum number of signal/background events to
  survive after final cuts, variables/processes to be used for a
  particular optimization run, number of optimizations inside one run
  to ensure that optimization converges/finds not just a local
  maximum(s), but a global one as well (in case of a complicated phase
  space);

\item to automatically verify stability of results.

\end{itemize}

This paper has the following structure: 

\begin{itemize}

\item Section 2 describes details of a ``toy''-study example,

\item Section 3 shows a simple example of a ``classical'', eye-balling
approach analysis for cuts optimization,

\item Section 4 gives details on a GARCON, GA approach to cuts
optimization and contains a comparison of these two approaches,

\item the following section is a detailed how-to user's manual.

\end{itemize}

The chosen ``toy-study'' is on purpose a simple Monte Carlo (MC)
analysis to illustrate GARCON functionality in a clear and transparent
way. Much more sophisticated use-cases of the program can be found
elsewhere~\cite{UF_SUSY,UF_h4mu,SINGLETOP}.

\section{LM6 with PYTHIA: a Toy Study}
\label{sec:pre-sel}
\label{section:simulation}

We are working in the framework of mSUGRA model \cite{msugra} which is
derived from more general MSSM ~\cite{MSSM} model using constrains
inspired by the super-gravity unification.  In case of mSUGRA, the
number of independent MSSM parameters is reduced to just five. For our
illustration we selected a point in mSUGRA parameter space with the
following values of mSUGRA parameters:

\begin{itemize}

\item the universal gaugino mass $m_{1/2} = 400$~GeV,

\item the scalar mass $m_0 = 85$~GeV,

\item the trilinear soft supersymmetry-breaking parameter $A_0 = 0$,

\item the ratio of Higgs vacuum expectation values, $\tan\beta = 10$,

\item sign of Higgsino mixing parameter, $sign(\mu) > 0$.

\end{itemize}

Characteristic qualities of SUSY events, following from a
consideration of signal Feynman diagrams are: large MET (mainly due to
massive stable SUSY particles, LSP) and large jet E$_{\rm T}$s (due to heavy
SUSY particles cascade decays). 

Background processes considered in this study are QCD, W/Z+jets,
double weak-boson production and $t\bar{t}$.

The main generation tool is PYTHIA 6.227 \cite{PYTHIA}. In addition,
ISASUGRA, part of ISAJET 7.69 \cite{ISAJET} is linked to PYTHIA to
provide mSUGRA masses, couplings and branchings for the signal
simulation.

All simulations and analysis is done for an integrated luminosity of
$10fb^{-1}$.

\subsection{Parameters of generation}

PYTHIA parameters for all generated processes are those for underlying
events. These parameters are specially tuned for LHC and used by both
ATLAS and CMS collaborations, defining underlying event physics can be
found elsewhere~\cite{PAOLO}. The simulation of the processes with
large cross-sections is performed in certain intervals of $\hat{\rm
p}_{\rm T}$. The list of simulated processes and their main
characteristics are listed in Table 1.

\begin{table}[htb]
\begin{center}
\caption{Data samples and their parameters}
\label{tab:QCD}
\vspace*{3mm}
    \renewcommand{\arraystretch}{1.5}
    \setlength{\tabcolsep}{2mm}
\begin{tabular}{|c|c|c|c|c|c|c|} \hline
~ & PYTHIA Process id  & Process & $\hat{\rm p}_{\rm T}$ (GeV/c) & Cross section$^{*}$ (pb)
& N$_{\rm generated}$ &   N$_{\rm expected}$ / N$_{\rm generated}$ 
 \\ \hline \hline 
1 & 39  & mSUGRA  &  no limits & 4 & 10$^{6}$ & 4$\cdot$10$^{-2}$ \\
2 & 16,31 & W + jet & 20-50  & 3.1$\cdot$10$^{4}$ & 9$\cdot$10$^{6}$ & 35.4 \\
3 &  -"-  &  -"-   & 50-100 & 7.9$\cdot$10$^{3}$ & 9$\cdot$10$^{6}$ &  8.8 \\
4 &  -"-  &  -"-   & 100-200& 1.5$\cdot$10$^{3}$ &7.02$\cdot$10$^{6}$& 2.1 \\
5 &  -"-  &  -"-   & 200-400& 1.4$\cdot$10$^{2}$ &1.44$\cdot$10$^{6}$ & 0.98\\
6 &  -"-  &  -"-   & $>$ 400  & 8.3              &8.5$\cdot$10$^{4}$ & 0.98 \\
7 & 15,30 & Z + jet & 20-50  & 1.2$\cdot$10$^{4}$ & 3$\cdot$10$^{6}$ & 38.2 \\
8 &  -"-  &  -"-    & 50-100 & 3.0$\cdot$10$^{3}$ & 3$\cdot$10$^{6}$ &  9.9 \\
9 &  -"-  &  -"-    &100-200 & 6.0$\cdot$10$^{2}$ & 3$\cdot$10$^{6}$ &  2.0 \\
10&  -"-  &  -"-    & $>$ 200& 6.4$\cdot$10$^{1}$ & 3$\cdot$10$^{6}$ &  0.21\\
11& 81,82 & $t\bar{t}$ & no limits & 840 & 8.0$\cdot$10$^{6}$  & 0.95 \\
12& 22,23,25& ZZ,WZ,WW &  -"-      & 1.4       & 1.0$\cdot$10$^{6}$  & 1.4$\cdot$10$^{-2}$ \\
13&       & QCD     &200-400 & 6.1$\cdot$10$^{4}$ & 5$\cdot$10$^{7}$ & 12.3 \\
14&       & -"-     &400-800 & 2.1$\cdot$10$^{3}$ & 5$\cdot$10$^{6}$ &  4.2 \\
15&       & -"-     & $>$ 800& 4.8$\cdot$10$^{1}$ &1.5$\cdot$10$^{5}$ & 0.32\\
\hline
\hline
\end{tabular} \\ ~ \\
\hspace*{-30mm} $^{*}$ For $t\bar{t}$, the NLO cross-section is assumed~\cite{TTCS}, for all 
other processes PYTHIA (LO) cross sections are taken.
         \\ ~ \\
\end{center}
\end{table}

\subsection{Variables and preselection}
\label{sec:variables}

Several variables characterizing the event were stored in the ASCII
files:

\begin{itemize}  

\item number of muons ($N_{\mu}$),

\item the highest muon p$_{\rm T}$ ($p^1_{\rm T}$),

\item isolation parameter for the highest p$_{\rm T}$ muon\footnote{${\rm ISOL
   = \sum{p^i_{\rm T}} }$ (p$_{\rm T}$ with respect to the beam direction) should be
   less or equal to 0, 0, 1, 2 GeV for the four muons when the muons
   are sorted by the ISOL parameter. The sum runs over only charged
   particle tracks with p$_{\rm T}$ greater then 0.8 GeV and inside a cone of
   radius ${\rm R = \sqrt{(\Delta\phi)^2 + (\Delta\eta)^2}=0.3 }$ in
   the azimuth-pseudorapidity space. A p$_{\rm T}$ threshold of 0.8 GeV roughly
   corresponds to the p$_{\rm T}$ for which tracks start looping inside the CMS
   Tracker. Muon tracks are not included in the calculation of the
   ISOL parameter} ($ISOL^1_{\mu}$),

\item number of jets with p$_{\rm T}$ $>$ 40 GeV ($N_j$),

\item E$_{\rm T}$ of the highest jet
  E$_{\rm T}$ ($E^1_{\rm T}$),

\item E$_{\rm T}$ of the third highest jet ($E^3_{\rm T}$),

\item missing transverse energy (E$_{\rm T}^{miss}$),

\item azimuthal angle between the highest-p$_{\rm T}$ muon and E$_{\rm
T}^{miss}$ (if any) ($\Delta\phi(\mu^1,E_{\rm T}^{miss})$),

\item azimuthal angle between the highest-E$_{\rm T}$ jet and E$_{\rm
T}^{miss}$ ($\Delta\phi(jet^1,E_{\rm T}^{miss})$),

\item circularity - {\it Circ} = 2 $\cdot$
  min($\lambda_{1},\lambda_{2}$) / ($\lambda_{1} + \lambda_{2}$),
  where $\lambda_{1},\lambda_{2}$ are eigenvalues of a simple matrix
  C$^{\alpha,\beta}$ = $\Sigma {\rm E}_{i}^{\alpha}{\rm
  E}_{i}^{\beta}$, where $\Sigma$ means sum over energies of all
  objects (leptons, jets, missing energy) and $\alpha,\beta$ = 1,2
  correspond to x and y components.  In case of back-to-back di-jets
  {\it Circ} is close to 0, while in case of multi-jet topology {\it
  Circ} tend to be closer to 1.

\end{itemize}

Jets are reconstructed using a cone algorithm with merging-splitting
of overlapping clusters.  In order to reduce the number of events in
the data files, a minimal E$_{\rm T}^{miss}$ cut of 50 GeV is applied
at generator level, which is known to be non-biasing, as typical
off-line cuts on E$_{\rm T}^{miss}$ are significantly higher
\cite{CMS_reach}. Another preselections include the requirement to
have at least two jets above 40 GeV in every event and a cut on the
leading jet {E$_{\rm T}$ in the event to be above 200~GeV. The latter
results from the fact that it doesn't look possible to simulate an
appropriate number of QCD events with $\hat{\rm p}_{\rm T}$ $<$
200~GeV/c

\subsection{Significance estimator}

The S$_{c12}$ significance estimator~\cite{s12} was used for
optimization: $S_{c12} = 2 \cdot (\sqrt{B+S}-\sqrt{B})$, where B - is
a number of all the background events after cuts, and S - is a number
of signal events after cuts. Results are presented also in terms of
$S_{cL} = \sqrt{2 \cdot (S+B) \cdot log(1.0+S/B)- 2 \cdot S}$ which
follows true Poisson probability for small number of events better than
$S_{c12}$, is shown in Ref.~\cite{UF_h4mu}.

\subsection{Splitting statistics in two parts}

We divided statistics in two parts: to perform cuts optimization on
one of them and then to verify stability of results on the other. It's
especially important for the analyses with limited statistics: in such
cases one risks to optimize cuts around a statistical fluke of a
signal over backgrounds significance. ``Blind experiment''
verification approach allows to exclude such unstable cases.

\section{Classical Search }
\label{section:calssical}

\subsection{Distributions and eye-balling search for cuts}
 
Figures \ref{fig:njets} -- \ref{fig:circ} show some of the simulated
data distributions which are used in the current analysis. Solid lines
denote the SUSY signal, while dashed lines - the sum of the SM
background distributions. 

\begin{figure}
  \begin{minipage}{0.4\columnwidth}
\resizebox{7cm}{!}{\includegraphics{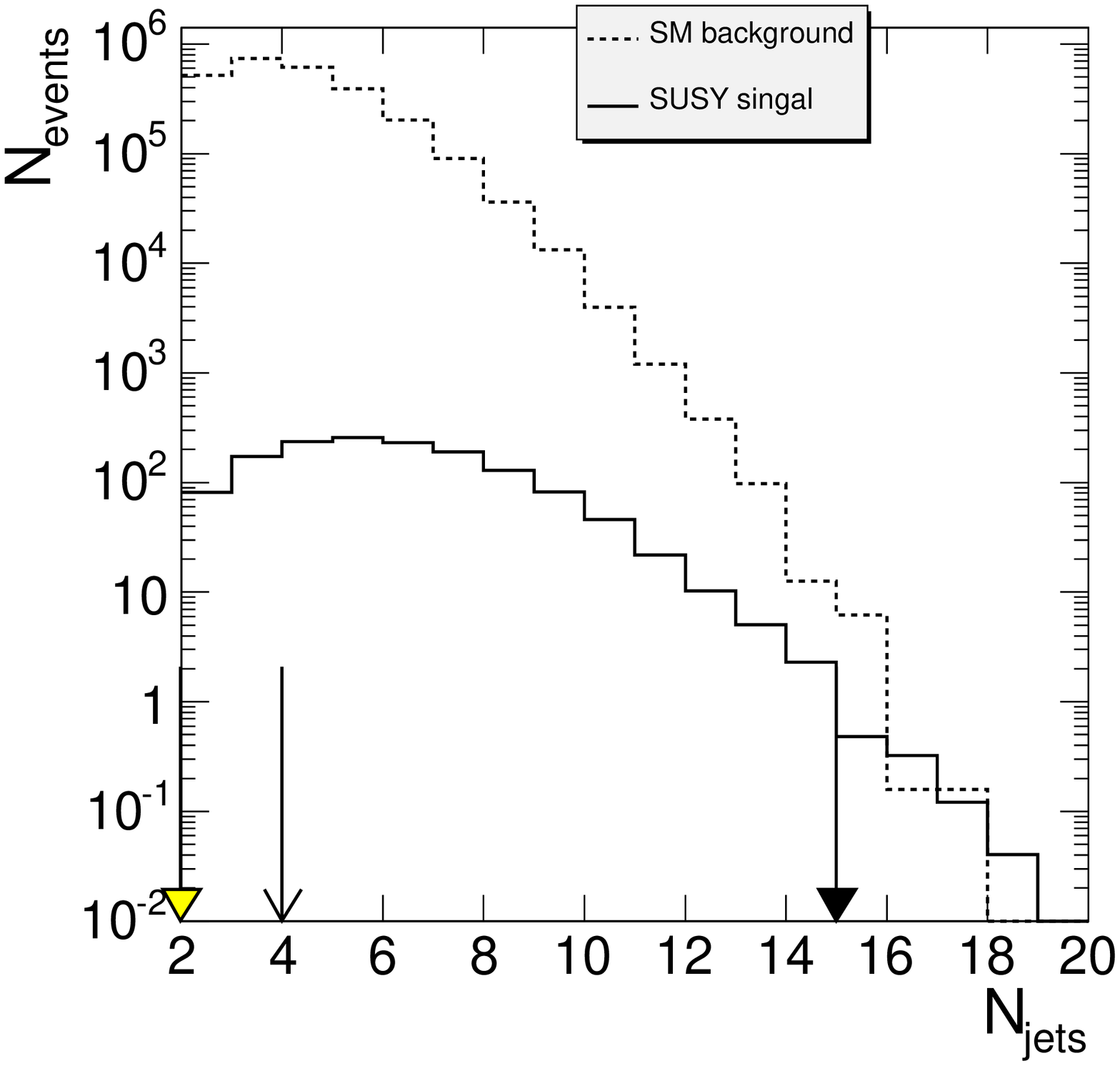}}
    \caption{Number of jets with E$_{\rm T}$ $>$ 40 GeV. Solid lines
denote the SUSY signal, while dashed lines - the sum of the SM
background distributions. Empty arrow is for classical analysis cut
choice, filled colored arrows (black and gray/yellow) are GARCON
optimized cuts (values for verification step).}
    \label{fig:njets}
  \end{minipage}
  \hfill
  \begin{minipage}{0.4\columnwidth}
    \resizebox{7cm}{!}{\includegraphics{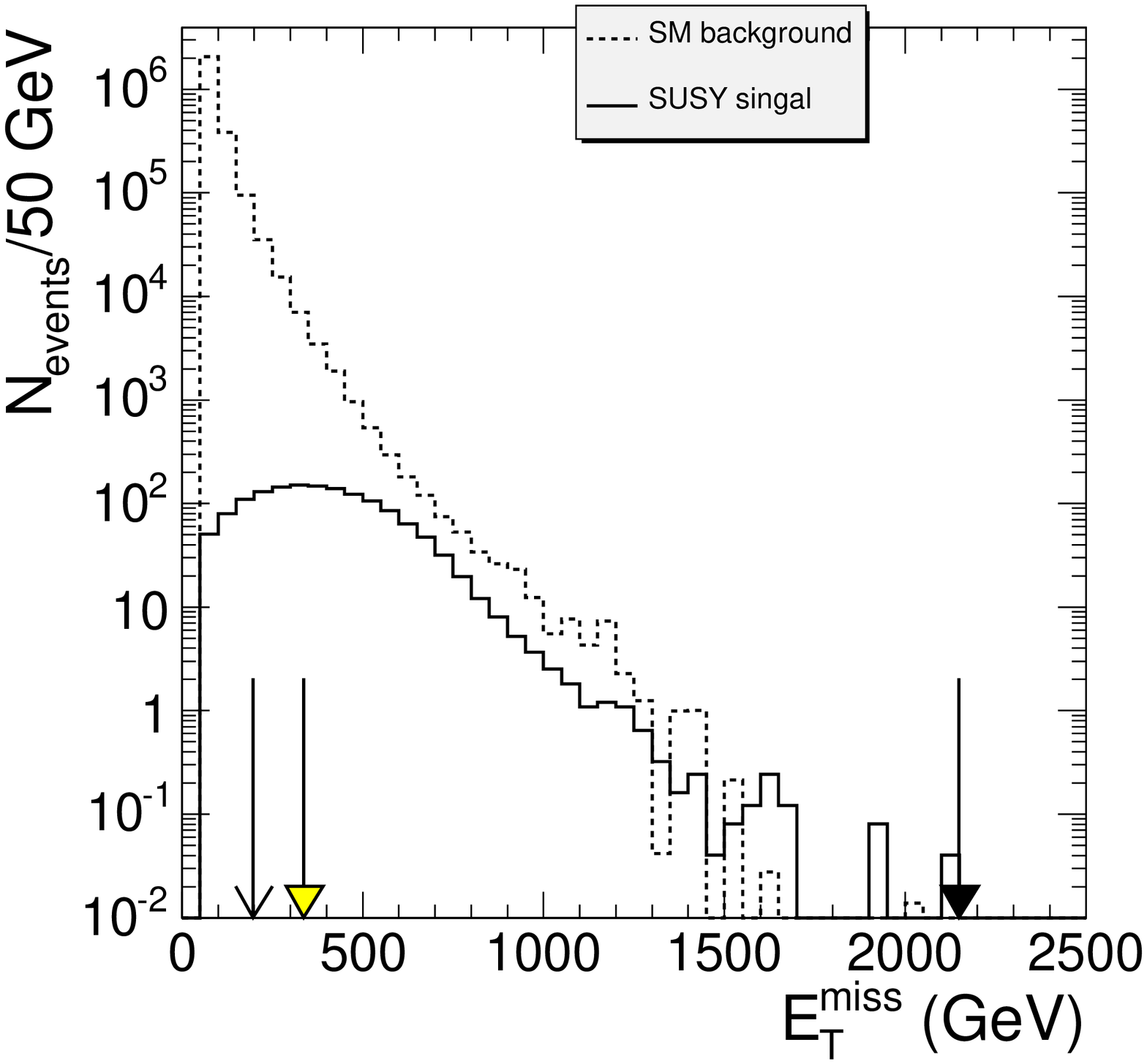}}
    \caption{Distribution of missing transverse energy. The same
    notations as for Fig.~\ref{fig:njets}.}
    \label{fig:met}
  \end{minipage}
\end{figure}

\begin{figure}
  \begin{minipage}{0.4\columnwidth}
\resizebox{7cm}{!}{\includegraphics{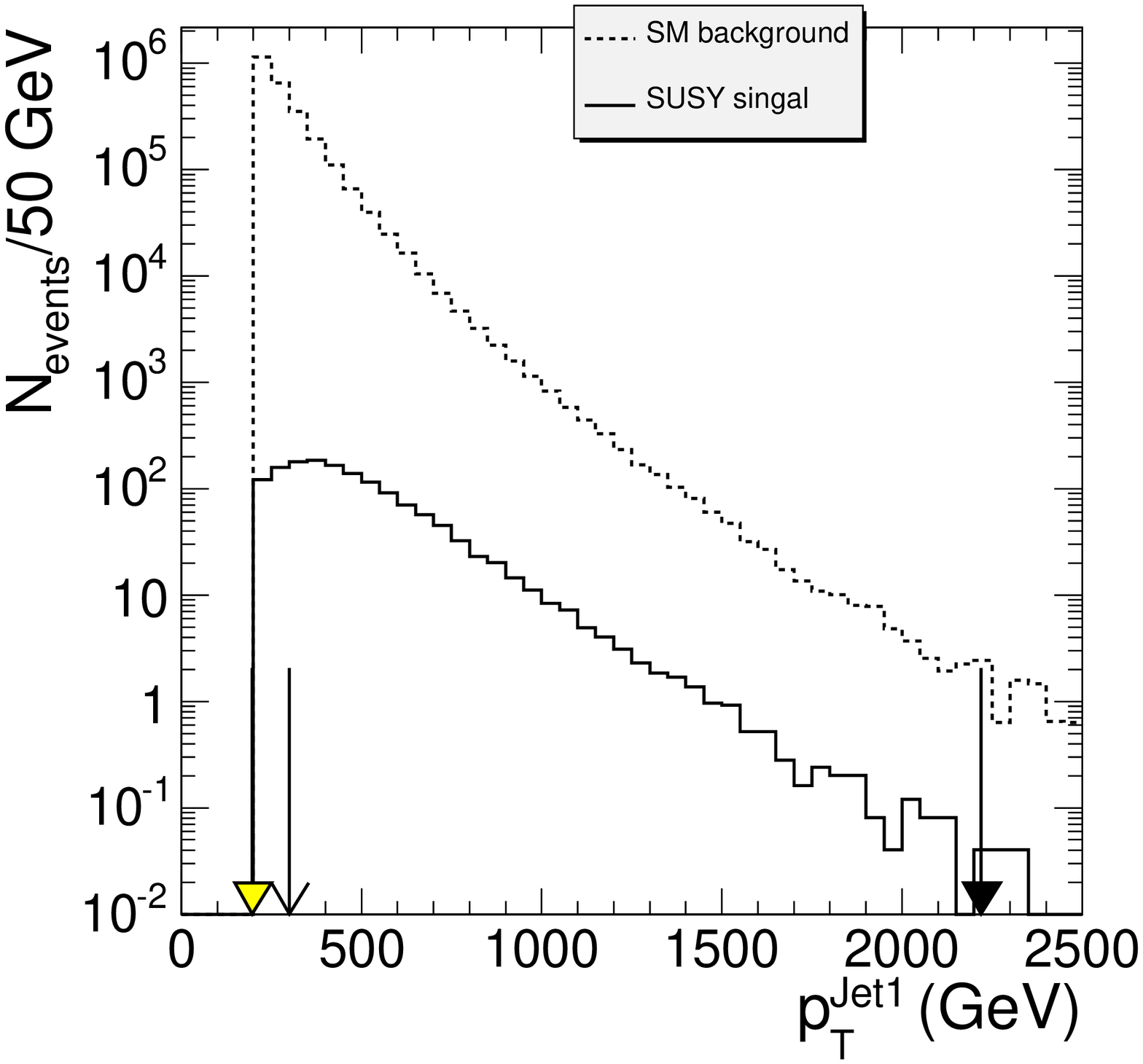}}
    \caption{Transverse energy of the hardest-E$_{\rm T}$ jet. The
    same notations as for Fig.~\ref{fig:njets}.}
    \label{fig:etjet1}
  \end{minipage}
  \hfill
  \begin{minipage}{0.4\columnwidth}
    \resizebox{7cm}{!}{\includegraphics{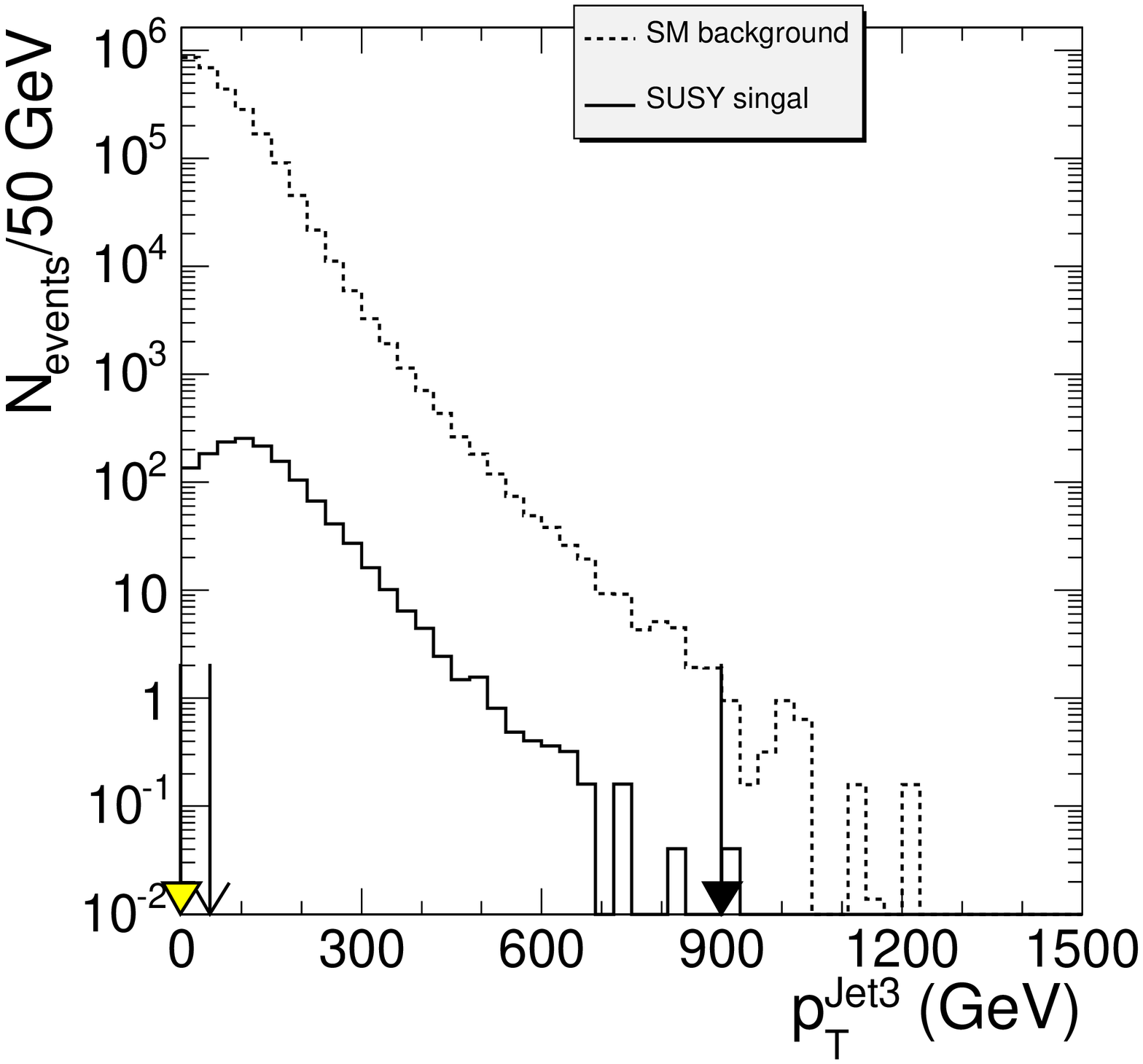}}
    \caption{Transverse energy of the third-hardest-E$_{\rm T}$
jet. The same notations as for Fig.~\ref{fig:njets}.}
    \label{fig:etjet3}
  \end{minipage}
\end{figure}

\begin{figure}
  \begin{minipage}{0.4\columnwidth}
\resizebox{7cm}{!}{\includegraphics{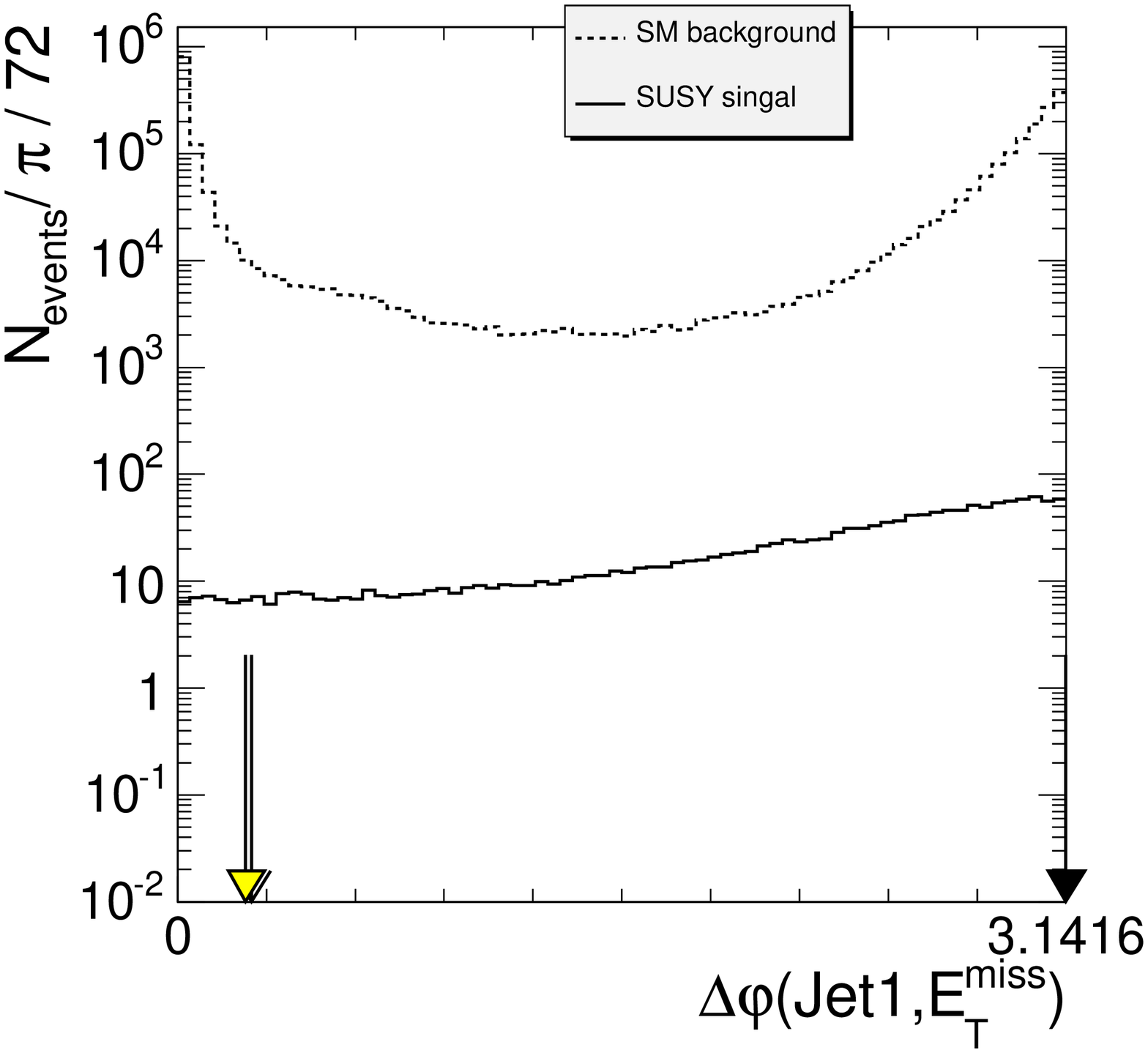}}
    \caption{Azimital angle distance between leading jet and
transverse missing energy. The same notations as for
Fig.~\ref{fig:njets}.}
    \label{fig:dphi}
  \end{minipage}
  \hfill
  \begin{minipage}{0.4\columnwidth}
    \resizebox{7cm}{!}{\includegraphics{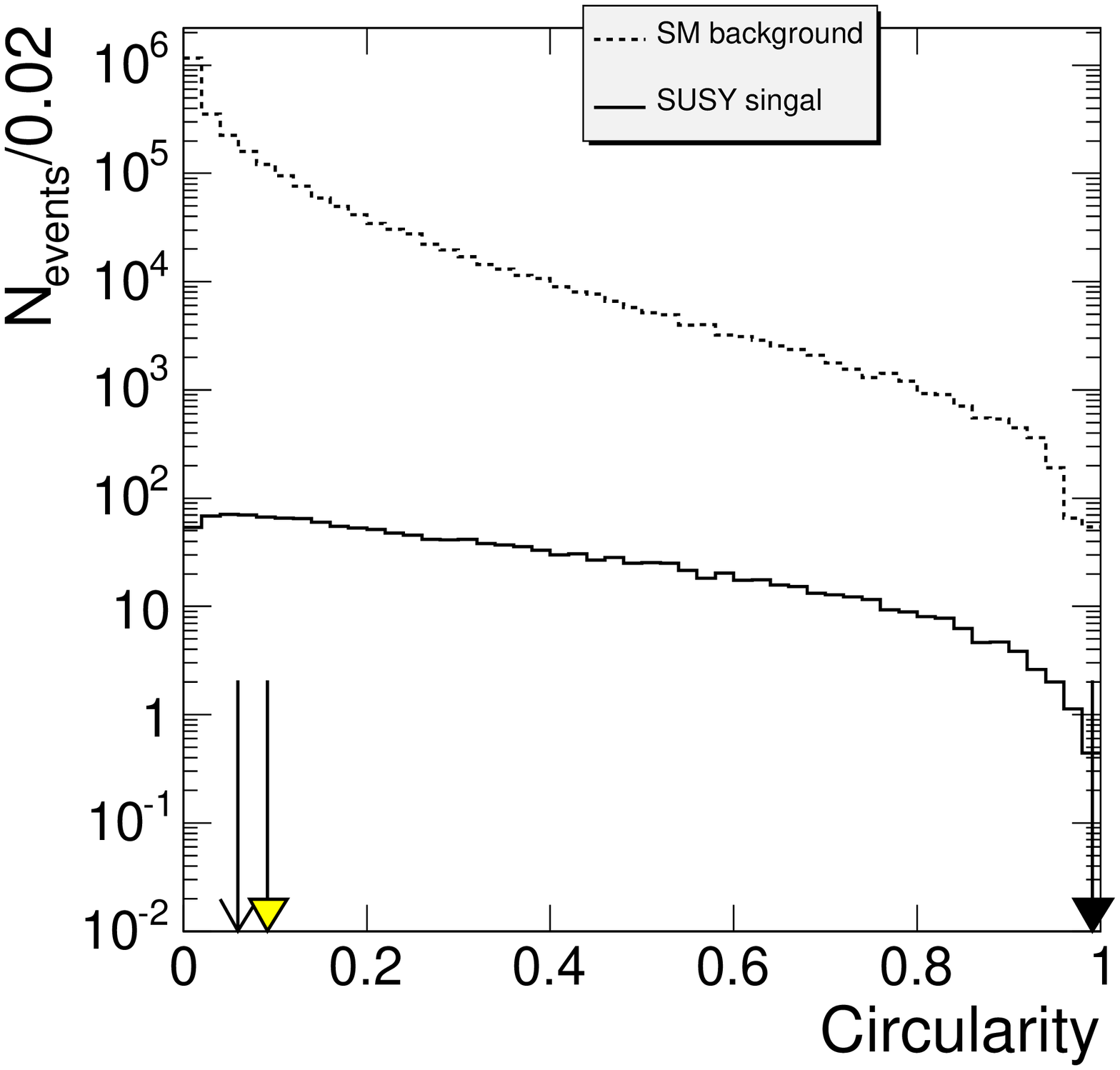}}
    \caption{Distribution of the circularity. The same notations
    as for Fig.~\ref{fig:njets}.}
    \label{fig:circ}
  \end{minipage}
\end{figure}

\vspace{10mm}

\section{GARCON Analysis}

\subsection{Evolution algorithms}

GARCON, GA-based programs in general, exploits evolution-kind
algorithms and uses evolution-like terms:

\begin{itemize}

\item {\it Individual} - is a set of qualities, which are to be
  optimized in a particular {\it environment} or set of
  requirements. In HEP analysis case, an Individual is a set of lower and
  upper rectangular cut values for each of variables under
  study/optimization.

\item {\it Environment} or set of requirements of evolutionary process
  in HEP analysis case is a {\it Quality Function} (QF) used for
  optimization of individuals. Significance of a signal over
  background is another widely used term in HEP community for QF. The
  higher QF value the better is an Individual. For a HEP analysis
  Quality Function may be as simple as $S/\sqrt{B}$, where S is a
  number of signal events and B is a total number of background events
  after cuts, or almost of any degree of complexity, including
  systematic uncertainties on different backgrounds,
  etc~\footnote{GARCON allows user defined QF}.

\item A given number of individuals constitute a {\it Community},
  which is involved in evolution process.

\item Each individual involved in the {\it evolution}, i.e. in
  breeding with a possibility of mutation of new individuals, death,
  etc. The higher is the QF of a particular individual, the more
  chances this individual has to participate in breeding of new
  individuals and the longer it lives (participates in more breeding
  cycles, etc.), thus improving community as a whole.

\item {\it Breeding} in HEP analysis example is a producing of a new
  individual with qualities taken in a defined way from two {\it
  ``parent''} individuals.

\item {\it Death} of an individual happens, when it passes over an age
  limit for it's quality: the bigger it's quality, the longer it
  lives.

\item {\it Cataclysmic Updates} may happen in evolution after a long
  period of {\it stagnation} in evolution, at this time the whole
  community gets renewed and gets another chance to evolve to even
  better quality level. In HEP analysis case it corresponds to a
  chance to find another local and ultimately a global maximum in
  terms of quality function. Obviously, the more complicated phase
  space of cut variables is used, the more chances exist that there are
  several local maximums in quality function optimization.

\item There are some other algorithms involved into GAs. For example
  {\it mutation} of a new individual. In this case, ``new-born''
  individual has not just qualities of its ``parents'', but also some
  variations, which in terms of HEP analysis example helps evolution
  to find a global maximum, with less chances to fall into a local
  one. There are also random creation mechanisms serving the same
  purpose, etc.

\end{itemize}

\subsection{Input for GARCON}

GARCON uses the same input information as a classical analysis: arrays
of variable values, see Appen.~\ref{sec:input}, the same what is
needed to perform a classical eye-balling cut optimization.

Details on chosen variables are given in Sec.~\ref{sec:variables}.

\subsection{Optimization}

Each cycle/``year'' of evolution includes a community update, that is
breeding process, possible mutation of new individuals, quality and
age calculations for each individuals, death of worst and too old
individuals, etc.

As described above the better is an individual QF, the longer it lives
and hence the more chances it has to produce new individs, improving
quality of a community as a whole and the very best individ quality as
a final goal. This very best individ or the very best set of min/max
cut variable values, which corresponds to the best achievable quality
function (significance of signal over background) is a final goal and
final output of the GARCON optimization step: rectangular cut values
recommended by the optimization procedure.

Figures~\ref{fig:dynMET}, \ref{fig:dynCIRC}, \ref{fig:dynS12} and
\ref{fig:dynTIME} show dynamics/evolution of the $S_{c12}$ quality
function, dynamics on MET and circularity cut variable values and
amount of time used for optimization.

Typical optimization procedure with GARCON takes from a few seconds to
several hours depending on the amount of statistics and additional
requirements like minimal number of events to survive after all cuts,
etc. As one can see from Fig.~\ref{fig:dynTIME} results close to the
best are already achieved before the first cataclysmic update, which
happend at year $<50$ and required less than 3.5 hours of CPU time for
10 variables (Sec.~\ref{sec:variables}) or 20 optimized parameters
with precision on each 2.5\% and about $4\cdot 10^5$ generated events
on input (after pre-selection, see Sec.~\ref{sec:variables}).

Optimized values for all the cut parameters are listed in
Table~\ref{tab:cuts}. Results in terms of chosen significance estimator
as well as signal to background number of events ratio, final event
numbers are listed in Tab.~\ref{tab:results}. Cuts are also
illustrated on cut parameter distribution
in Figs.~\ref{fig:njets}-\ref{fig:circ}.

\begin{table}[htb!]
  \begin{center}
    \caption{Min and max values for cut parameters. Cut values for the
    classical analysis are the same. Cut values for GARCON
    verification are rounded off in comparison to those we have from
    optimization to reflect resolution effects and possible
    lower/upper limits. \label{tab:cuts}}
    \begin{tabular}{|c|c|c|c|} \hline
      cut parameter & classical & GARCON optimization & GARCON
      verification \\ \hline \hline
      $N_{mu}$                & 0-inf      & 0-5          & 0-5        \\ \hline
      $p^1_{\rm T}$, GeV         & 0-inf      & 0-1020       & 0-inf      \\ \hline
      $ISOL^1_{mu}$, GeV       & 0-inf      & 0-1080       & 0-inf      \\ \hline
      $N_j$                 & 4-16       & 2-16         & 2-16       \\ \hline
      $E^1_{\rm T}$, GeV          & 300-inf    & 200-2220     & 200-inf    \\ \hline
      $E^3_{\rm T}$, GeV          & 50-inf     & 0-901        & 0-inf      \\ \hline
      E$_{\rm T}^{miss}$, GeV           & 200-inf    & 342-2150     & 340-inf    \\ \hline
      $\Delta\phi(\mu^1,E_{\rm T}^{miss})$, rad & 0- $\pi$  & 0.297-$\pi$     & 0.297-$\pi$   \\ \hline
      $\Delta\phi(jet^1,E_{\rm T}^{miss})$, rad  & 0.262-$\pi$ & 0.245-$\pi$ & 0.245-$\pi$ \\ \hline
      circularity               & 0.06-1     & 0.0924-0.993 & 0.0924-1   \\ \hline
    \end{tabular}
  \end{center}  
  
\end{table}

Analysing cut values and their distributions
(Figs.~\ref{fig:njets}-\ref{fig:circ}) one can see that some variables
after GARCON optimization converge to the limits of a particular
distribution. From the technical point of view the reason for this is
because GARCON works only with input values and doesn't have plus or
minus infinity e.g. at its disposal. From the practical point of view,
it means that min or max cut on a particular variable or the whole
variable is not useful in comparison to other variables in terms of
improving signal to background significance and GARCON shows it. As an
example we can consider $E^1_{\rm T}$ and $E^3_{\rm T}$ before and
after all cuts (except the cut on $E^1_{\rm T}$ or $E^3_{\rm T}$
correspondingly), the examples of variables for which GARCON and
classical cut values are different: compare Figs.~\ref{fig:etjet1} and
\ref{fig:etjet3} for distributions before and
Figs.~\ref{fig:etjet1after} and \ref{fig:etjet3after} - after the cuts
applied.

\subsection{Verification}

As mentioned earliler, the available MC statistics was divided in two
parts. The second part is used for a ``blind'' analysis or results
stability verification.

After we got the cut values from optimization step, we round them off
to the level of expected precision~\footnote{Expected precision, which
includes detector resolution, of course is different for different
parameters (muon $p_{\rm T}$, jet $E_{\rm T}$) and different HEP
experiments.} for each parameter (see Tab.~\ref{tab:cuts}) and apply
them to the second half of the statistics.

Results are shown in Tab.~\ref{tab:results}. One can see that results
are stable\footnote{NOTE: in case there are zero generated events left
after final cuts we use $0 \pm 1$ generated events, taking slightly pessimistic
estimation for MC statistical error and hence corresponding number of expected
events: $0 \pm one-generated-event-weight$.}.

\begin{table}[htb!]
  \begin{center}
    \caption{Final results comparison for classical and GARCON (for
    optimization and verification steps) approaches in terms of
    $S_{c12}$ and $S_{cL}$ significance estimators as well as ratio of
    final number of signal to total background events and those
    numbers of events with MC statistical error
    included. \label{tab:results}} 
    \begin{tabular}{|c||c|c||c|c|} \hline
      parameter & classical optimization  & classical
      verification & GARCON optimization & GARCON verification \\
      \hline
      $S_{c12}$           & 8.1            & 8.0            & 15.3           & 14.7   \\ \hline
      $S_{cL}$            & 8.1            & 8.1            & 15.8           & 15.2   \\ \hline
      S/B                 & 0.102          & 0.102          & 0.506          & 0.469   \\ \hline
      $N_{signal}$        & 665 $\pm$ 7    & 663 $\pm$ 7    & 574 $\pm$  7   & 567 $\pm$ 7   \\ \hline
      $N_{background}$    & 6496 $\pm$ 160 & 6503 $\pm$ 160 & 1130 $\pm$ 121 & 1210 $\pm$ 121   \\ \hline
    \end{tabular}
  \end{center}  
  
\end{table}

\begin{figure}
  \begin{minipage}{0.4\columnwidth}
\resizebox{7cm}{!}{\includegraphics{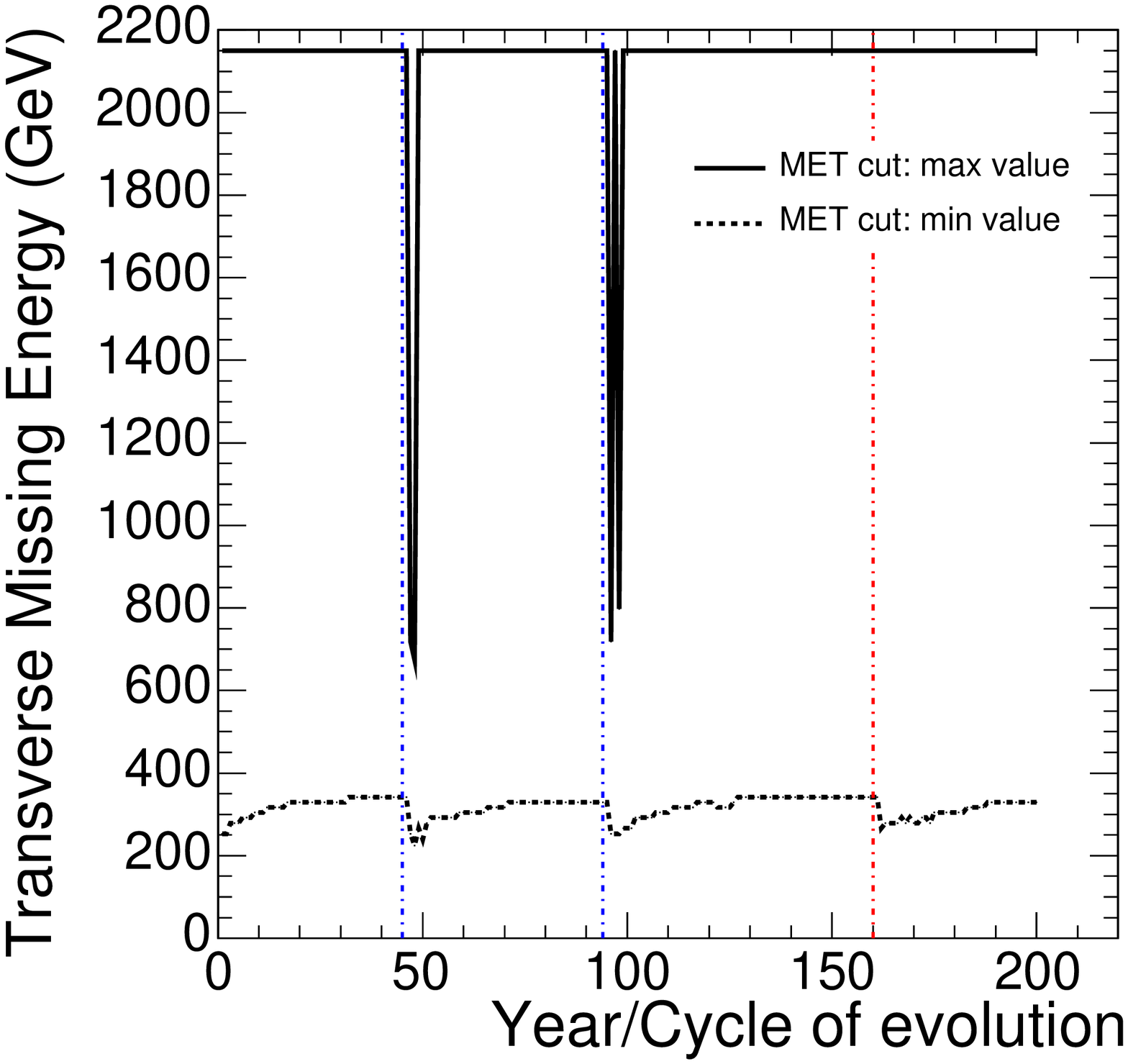}}
    \caption{Evolution of the cuts on MET. Upper and lower
    curves are for min and max cut values on the variable. Vertical
    dotted-dashed lines show cataclysmic update times, the right one
    corresponds to a cataclysmic update after the best result
    achieved.}
    \label{fig:dynMET}
  \end{minipage}
  \hfill
  \begin{minipage}{0.4\columnwidth}
    \resizebox{7cm}{!}{\includegraphics{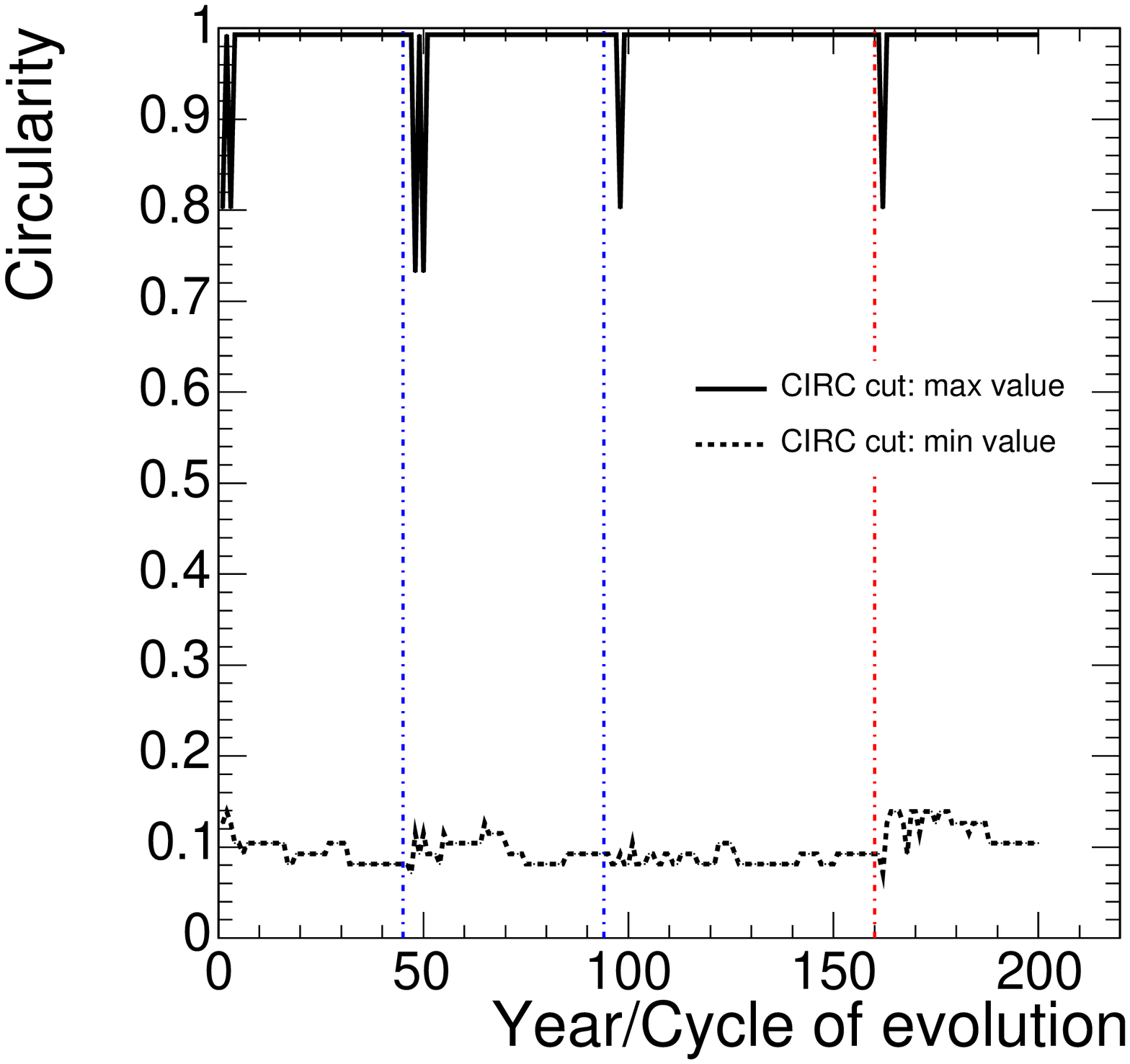}}
    \caption{Evolution of the cuts on CIRC. Notations are the
    same as for Fig.~\ref{fig:dynMET}.}
    \label{fig:dynCIRC}
  \end{minipage}
\end{figure}

\begin{figure}
  \begin{minipage}{0.4\columnwidth}
\resizebox{7cm}{!}{\includegraphics{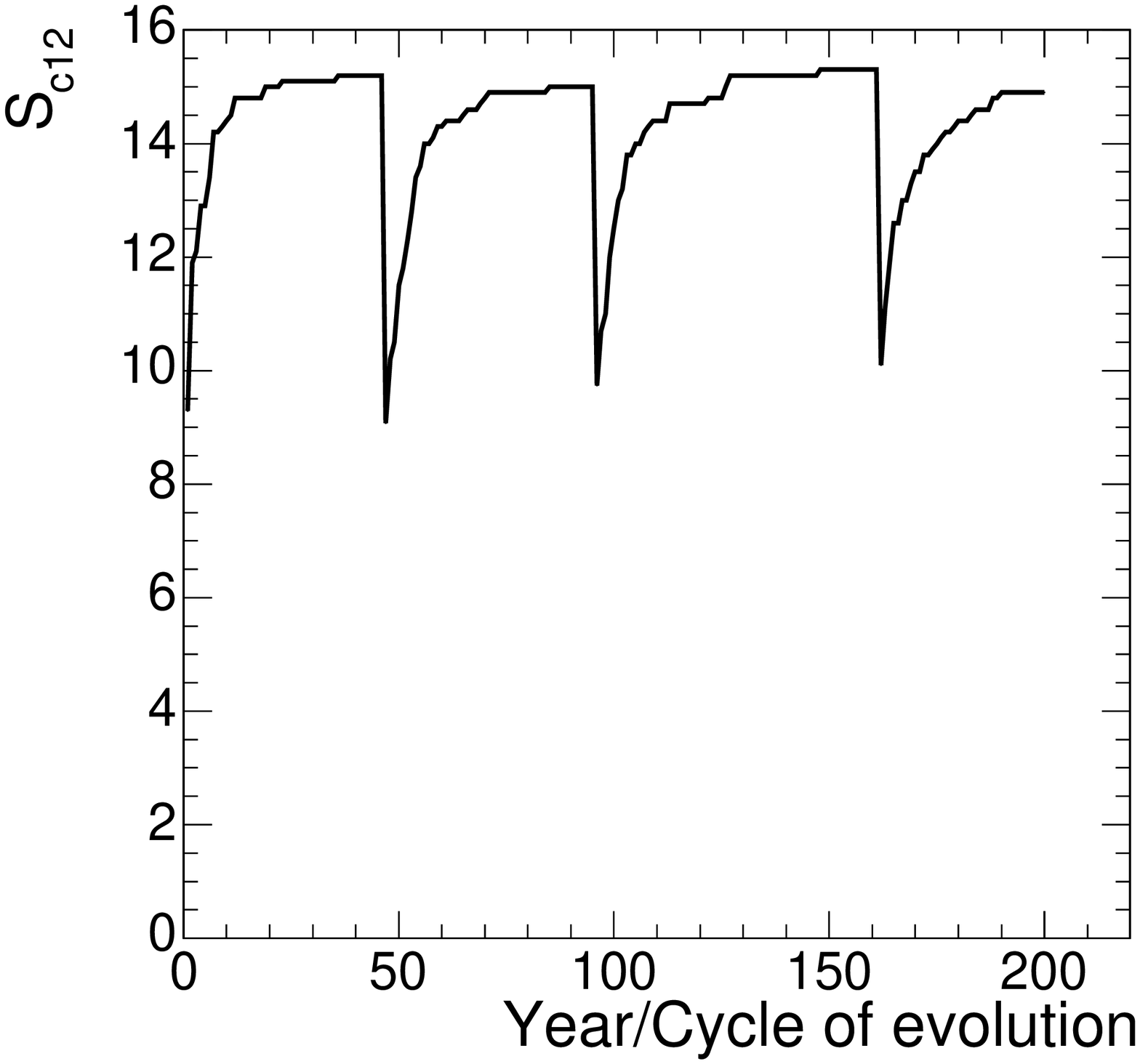}}
    \caption{Significance ($S_{c12}$) estimator value
    dynamics. Notations are the same as for Fig.~\ref{fig:dynMET}.}
    \label{fig:dynS12}
  \end{minipage}
  \hfill
  \begin{minipage}{0.4\columnwidth}
    \resizebox{7cm}{!}{\includegraphics{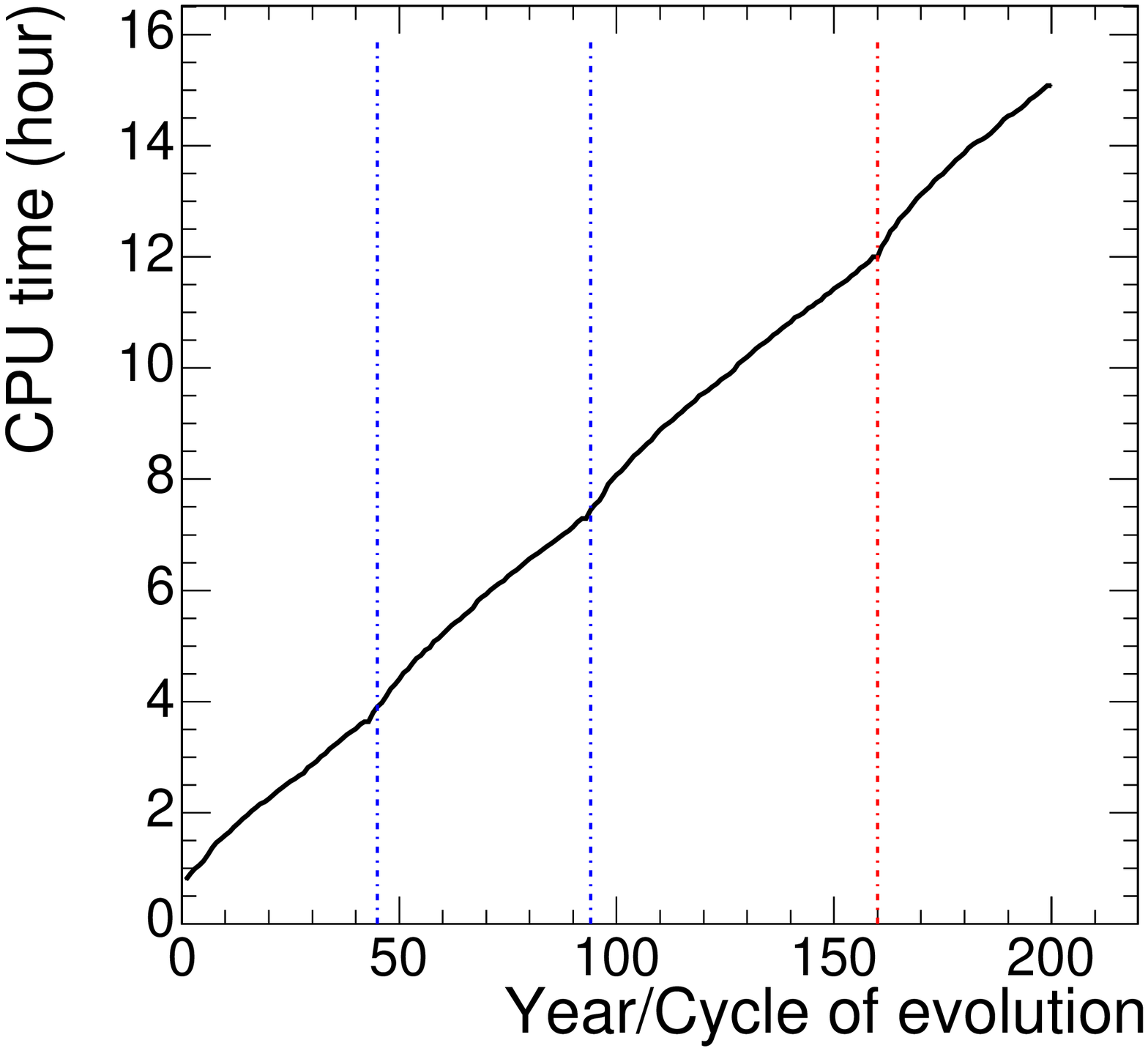}}
    \caption{Amount of time spent for evolution. Notations are the
    same as for Fig.~\ref{fig:dynMET}.}
    \label{fig:dynTIME}
  \end{minipage}
\end{figure}

\begin{figure}
  \begin{minipage}{0.4\columnwidth}
\resizebox{7cm}{!}{\includegraphics{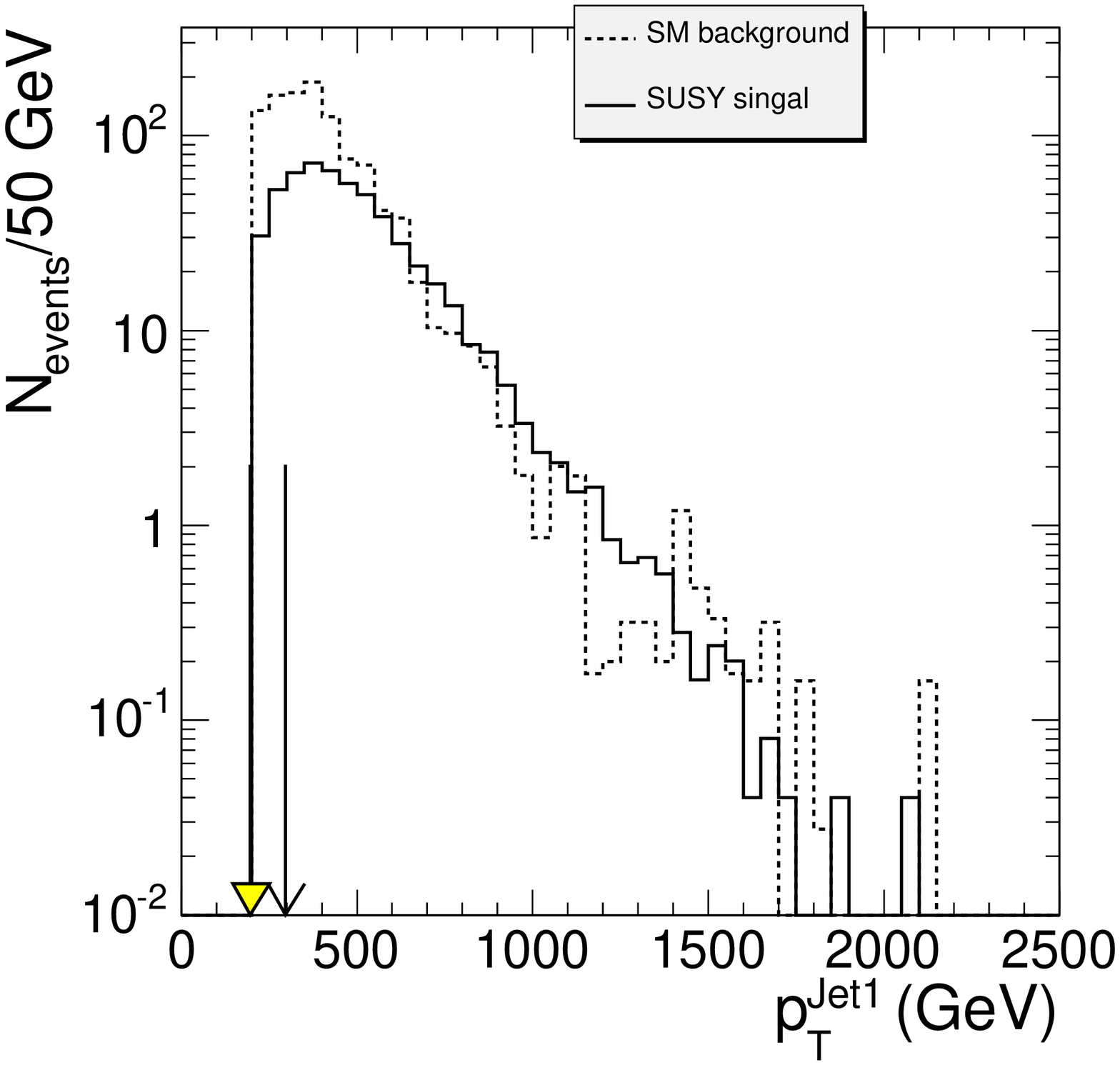}}
    \caption{Transverse energy of the hardest-E$_{\rm T}$ jet. The
    same notations as for Fig.~\ref{fig:njets}.}
    \label{fig:etjet1after}
  \end{minipage}
  \hfill
  \begin{minipage}{0.4\columnwidth}
    \resizebox{7cm}{!}{\includegraphics{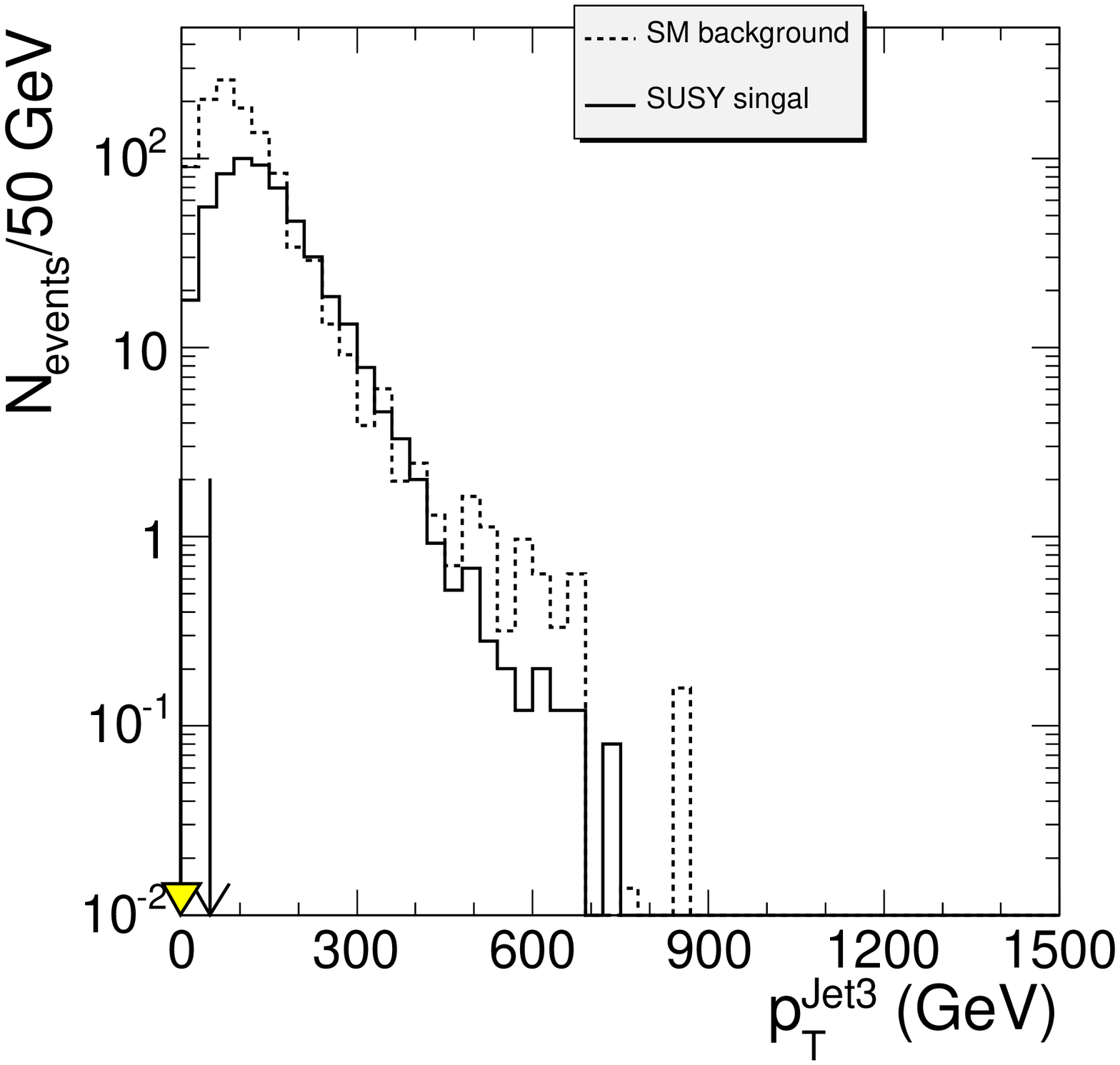}}
    \caption{Transverse energy of the third-hardest-E$_{\rm T}$
jet. The same notations as for Fig.~\ref{fig:njets}.}
    \label{fig:etjet3after}
  \end{minipage}
\end{figure}

\subsection{Comparison between classical and GARCON approaches}

Difference in performance in terms of significance (8 vs. 15) and
signal to background events number ratio (0.1 vs. 0.5) may not be a
typical gain when GARCON is used vs. a classical approach: classical
approach may be pretty sophisticated (as well as time dedicated to it
may be large). What is important to emphasize is that GARCON does
optimization and verification of results stability in an automatic
manner, not requiring any special treatment of either input data or
output results and does converge to virtually the best set of cuts in
typically hours time.

Different cases which show GARCON usage in much more complicated
analyses cases can be found elsewhere
\cite{UF_SUSY,UF_h4mu,SINGLETOP}.

\section{GARCON v2.0: User's Manual.}
\label{sec:manual}

\subsection{Introduction}

This is a user's manual on how to run GARCON, GA program v2.0 and use
its output in physics analyses. The program is designed for
rectangular cuts optimization (with same interpretation and usage of
cut values as you would expect from classical, eye-balling rectangular
cut values selection).

You may find useful the following presentations linked to the GARCON home
page~\cite{GARCONMAIN}:

\begin{itemize}

\item ``GARCON - Genetic Algorithm for Rectangular Cuts OptimizatioN'' -
  {\color{red} this is a how-to use the program talk}

\item ``Genetic Algorithm studies and comparisons using different significance estimators''

\item ``Genetic Algorithm and example application in a SUSY search''

\end{itemize}

\subsection{Installation}

For now to make user's and developers life easier code is available as
a static library pre-compiled at Scientific Linux PCs (CERN's Linux
Red Hat 9.0 version) it has a template for a user to define
optimization function (several widely-used functions described below
are also available). All you need is to get an archive from the
following web-page (look for ``Code'' link there) and do:

\begin{itemize}

\item go to: http://drozdets.home.cern.ch/drozdets/home/genetic/

\item download the most recent version of the program ({\color{blue}
garcon-2\_0.tar.gz}),

\item do {\color{blue} 'tar -xzf garcon-2\_0.tar.gz'} in any directory
  you would like to work with GA,

\item read and follow README file instructions on how to build
  executable and run GARCON.

\end{itemize}

You will find there the following files/directories:

\begin{itemize}

\item {\color{blue}lib/libgenetic.a} - is a GARCON library file.

\item {\color{blue}quality.cc} - C++ template file for you to define
  your own quality/significance function for optimization. (Look
  inside this file, it has two examples of user's functions with
  detailed comments: simple and detailed ones. You should be able to
  easily construct your own function in a similar way.)

\item{\color{blue}Makefile} - for making binary file: {\color{blue}garcon-ga}.

\item {\color{blue}data/} - a directory, where you can store your
  input data files (see example of their format below). 

\item {\color{blue}dataFiles.dat, initialization.dat,
  verification.dat, dataErrors.dat, variablesON\_OFF.dat} - files with
  input parameters (description is given below).

\end{itemize}

\subsection{How to run GA.}

Just use {\color{blue} './garcon-ga $>$ output.txt'} after you have
prepared an executable following instructions in README file and put
appropriate parameters into dataFiles.dat, initialization.dat,
verification.dat, dataErrors.dat files.

\subsection{Input data sample files format.}

Example is in: {\color{blue}data/example.dat}, see also
Appen.~\ref{sec:input}. {\color{red}NOTE: This file is just an example
of format, it is a shortened version of a real data file.}

First line is for a process name (one word). {\color{green}In example
it is: lm1.}

Second line is for weight per event in sample (usually weight
calculated as: $weight = \frac{\sigma \int L
\epsilon}{N_{events-in-sample}}$). Do not forget to correspondingly
increase weights when you divide your statistics into two parts, one
for optimization and one for verification steps. {\color{green}In
example it is: 0.324} One can also vary weight with {\it weightCoef}
parameter described in Sec.~\ref{sec:param}.

Third line is for variable/cut names (one word for each). In the example file
there are 12 of them: {\color{green}met njet et\_jet1 et\_jet2 eta\_lep1
eta\_lep2 deltaR cosTheta dphi\_l1met dphi\_j1met njet30 njet50}.

Forth and the following lines are for input data. Values of the above
listed variables for each event for a particular process. ({\color{red}One input
file per process! One line per event!}) See example in the file.

\subsection{Input parameters. File dataFiles.dat.}
\label{sec:data}

This file in each line has a PATH to a particular input data
sample. The paths may be relative (relative with respect to the
directory where you run GA) or absolute (always works).

Example is in: {\color{blue}dataFiles.dat} release file version.

{\color{red}NOTE: input file for signal is always the last one!}

You will need two such files with two lists of input files if you are
to perform optimization of cuts and verification of results stability.

\subsection{Input parameters. File initialization.dat.}
\label{sec:param} 

This file has input parameters for GA, see Appen.~\ref{sec:ini}.
\begin{itemize}

\item {\color{blue}internal, 0} - three first service parameters. Just
put them 0, they are not used in the public version and exist for
debugging purposes.

\item {\color{blue}Integer maxNumberFeatures} - the number of cut
  parameters used in your analyses (with values provided in input
  files of course). It equals to 12 in the described above input data
  sample files format example.

\item {\color{blue}Integer maxNumberProcesses} - should be equal to
  number of input files. It equals to 9 for the
  {\color{blue}dataFiles.dat} release file version.

\item {\color{blue}Integer maxNumberFeatureValues} - this says to GA
  how precise you want your cuts, how small step to use. For example
  if you put maxNumberFeatureValues equal 40, it means that cut step
  corresponds to 2.5\% (100/40) of events cut every other cut
  value. (Signal distributions are used to define cut values. Steps
  are not equal, they depend on events density for each distribution.)

\item {\color{blue}Integer initNumberPopulation, greater than 1} -
  this would be the number of different cut sets (Individuals)
  involved in evolution. Better to avoid settings too small ($<<30$)
  or too big ($>>500$). Default value of 100 is a reasonable choice.

\item {\color{blue}Integer maxNumberBestIndivids, greater than 0 and
  smaller than initNumberPopulation} - should be much smaller than
  initNumberPopulation. This is the number of the best cut sets. These
  are cut sets which get priority in GA iteration steps. (There is
  always one the very best cut set which is printed in all the
  details.) Default is 5 (for 100 initNumberPopulation).

\item {\color{blue}Integer ageLimit, greater than 0 and smaller than
  yearsForEvolution} - how long each cut set (Individual) will be
  involved in evolution iterations. ageLimit is also the limit of how
  long the very best cut set may not change before the whole
  population of cut sets will be forced to get new try (cataclysmic
  update). (This ``new try'' or critical update allows GA to try to
  find another maximum in case there are more than one local max for
  significance in a given parameter space.) Values between 10-50 are
  good to try. Default is 10.

\item {\color{blue}Real mutationFactor, from 0 to 1} -
  technical. Shows a degree of randomness in mutation process. Default
  is 0.8. Values between 0.01 and 1.00 can be tried.

\item {\color{blue}Integer yearsForEvolution, greater than 0 (less
  than 1000)} - number of iteration cycles for the whole
  optimization. Several hundred are OK. (This number should be several
  times bigger than ageLimit.)

\item {\color{blue}Integer optimFlag, 1 or 0} - 1, if you do
  optimization, 0 if you perform verification of results. First is to
  be performed on one part of the statistics you have in analysis and
  is for finding the best cuts set. The second is to verify stability
  of the results using output from the optimization step.

\item {\color{blue}Integer SignificanceChoice, 0, 1, 2, 3, 4, 5 or 6}:

\begin{itemize}

\item 0 is for $S_1 = S/\sqrt{B}$, 

\item 1 is for $S_2 = S/\sqrt{S+B}$, 

\item 2 is for $S_{c12} = 2 \cdot (\sqrt{B+S}-\sqrt{B})$,

\item 3 is for $S_{cL} = \sqrt{2 \cdot (S+B) \cdot log(1.0+S/B)- 2
  \cdot S}$,

\item 4 is for $S/B$, where B - is a number of all the background
  events after cuts, and S - is a number of signal events after cuts,

\item 5 and 6 are for user defined significance functions (5 is for a
  simple one and 6 allows user to access details for each event, see
  Appen.~\ref{sec:quality}).

\end{itemize}

Look at the ``Genetic Algorithm studies
  and comparisons using different significance estimators'' talk for a
  hint on a particular significance estimator stability and
  differences between them.

\item {\color{blue}Real minEventsCoefficientSignal and
  minEventsCoefficientBackground} - minimal number of events,
  which should survive after cuts, for background processes calculated
  as $minEventsCoefficientBackground \cdot
  \sqrt{\sum_{i=1}^{Nbackground-processes} weight_i^2}$ and
  $minEventsCoefficientSignal \cdot weight_{signal}$ for signal
  process. Default is 5. This parameter or final events number
  thresholds affect results stability as they do in a classical
  approach as well.

\item {\color{blue}Real weightCoef} - a re-weighting coefficient. For
  example if you have your samples prepared for 10$fb^{-1}$ and would
  like to see how results of optimization would change for other
  luminosities, for example for 100$fb^{-1}$, you can simply put
  weightCoef = 10. This parameter is also useful when you divide your
  statistics to two parts for optimization and verification and don't
  want to remember to change weights in all the data samples, you may
  just change weightCoef (if you divide statistics half-by-half, you
  need to multiply weight for every sample by 2, or put weightCoef=2
  to have calculations done for the same integrated luminosity). 

Example is in: {\color{blue}initialization.dat} release file version.

\end{itemize}

\subsection{Input parameters for verification. File verification.dat.}

First line is for number of cut sets to be verified. In example it is
4, see Appen.~\ref{sec:verif}. 

Each of the best cut sets is listed in an output file after
optimization step. The very best one for each iteration is printed in
all details with cut values, and there are some details on
maxNumberBestIndivids of best cut sets (cut values, age, etc). 

So, you may use the following procedure after optimization step is
done. Do: {\color{blue}'grep Calculated output.txt'}, you will get a
list of the best values of significances. You will likely see that
there were several attempts by GA to find the best optimization
(significance/quality increases, then stays stable, then starts over
again). So, you may find out looking at the output file what cut sets
({\color{blue}'Min Individ Feature Values' and 'Max Individ Feature
Values'} - upper and lower cut values) corresponds to a few maximums
you find in the 'grep Calculated output.txt' listing. Just cut and
paste them into verification.dat (two lines per cut set: 'Min Individ
Feature Values' and 'Max Individ Feature Values'). There are four such
pairs in example file.

Then run GA again, but with optimFlag set to 0. Better to do this on a
different part of the statistics (``blind'' experiment) and with cut
parameter values rounded off to levels of corresponding precisions to
avoid cuts ``overtuning''.

\subsection{Input parameters. File dataErrors.dat.}

This file has a line of ``penalty/priority'' factors for each process
in order as they are listed in dataFiles.dat, see
Appen.~\ref{sec:penalty}.

Example is in: {\color{blue}dataErrors.dat} release file version.

{\color{red}NOTE: input file for signal is always the last one!}

The idea of the ``penalty/priority'' factors is to apply a simple
estimation on systematic effects influences (alternative possibility
is to define a sophisticated user QF, see Sec.~\ref{sec:param} and
Appen.~\ref{sec:quality}). If the factor values are different from
0.0, then equations for Significance Functions described above will be
calculated with modified number of signal and background events: $S
\rightarrow S+kS \cdot S$, $B = \sum{B_i} \rightarrow B =
\sum(B_i+kB_i \cdot B_i)$, where $kS$ and $kB_i$ are introduced in
dataErrors.dat factors.

{\color{blue}HINT:} putting a factor equal to -1.0 will effectively switch off a
particular process. (-1.0 is the minimal value for a factor, upper
value is not limited.)

{\color{blue}HINT:} putting $kB_i >= 0$ and $kS <= 0$ will provide you
with a conservative estimation (opposite signs - with optimistic
values of Significance).

{\color{blue}HINT:} one may try these factors after optimization,
during ``verification step'', getting a feeling of different scenarios.

\subsection{Input parameters. File variablesON\_OFF.dat.}

This file has a line of ON/OFF switches for each parameter in input
files in order as they are listed in input data files, see
Appen.~\ref{sec:onoff}. This allows user to prepare input data files
with exhaustive list of possible cut parameters and then perform
optimization studies on different combinations of the parameters.

\subsection{How to use GA results.}

Basically cut sets (pairs of 'Min Individ Feature Values' and 'Max
Individ Feature Values') for each particular cut/parameter are similar
in use and interpretation as classical, eye-balling cuts one usually
looking for to discriminate between signal and background event. So,
once you've got GA results and verified their stability you may go on
with your analysis as after you find out a set of classical
rectangular cuts for your distributions.

{\color{blue}HINT:} the very best result is repeated at the end of the
output.

{\color{blue}HINT:} output contains details on dynamics of all the
significance estimators available (while performing optimization on
one of them).

{\color{blue}HINT:} you may want to use not the very best Individual
(cuts set), but for example a one corresponding to a local maximum
with worse performance, but better stability. As described above
(Sec.~\ref{sec:param}) one of the means of making results stable is to
ask for a particular number of events to survive. Obviously, if weight
per generated event for a particular process is something like 1000
expected ``real'' events and cuts kill all the events in MC sample,
one effectively has $0 \pm 1000$ events expected, which (sero survived
events) may be a statistically unstable result.

\section{Summary}

We presented GARCON program, illustrated its functionality on a simple
HEP analysis example, much more complicated examples described for
example in the CMS Physics Technical Design Report. The program
automatically performs rectangular cuts optimization and verification
for stability in a multi-dimensional phase space.

All-in-all it is a simple yet powerful ready-to-use publicly available
tool with flexible and transparent optimization and verification
parameters setup.

\section{Acknowledgments}

We would like to thank
A.~Giammanco, 
A.~Korytov and
A.~Sherstnev
for useful discussions.

\clearpage

\newpage

\appendix

\section{Input data file format.}
\label{sec:input}

A part of one input data example file is given below.

{\footnotesize
\begin{verbatim}
lm1
0.324
met njet et_jet1 et_jet2 eta_lep1 eta_lep2 deltaR cosTheta dphi_l1met dphi_j1met njet30 njet50
383.926 2 550.591 150.944 -0.800319 -1.32351 0.864812 0.86366  0.191271 2.90746 2 2
229.268 3 248.019 103.515 0.930838  -1.82199 3.82118 -0.883346 0.772977 2.64147 2 2
149.199 6 374.811 142.887 -1.5395   -1.73921 1.44724  0.876463 2.00948 1.74833 5 3
266.147 5 369.581 131.862 1.49267    1.98226 0.950259 0.949556 0.6018 2.96775 3 2
360.203 3 242.108 158.631 -1.63074  -1.54017 2.37681  0.733466 2.4858  2.4815 2 2
\end{verbatim}
}

\section{Initialization data format.}
\label{sec:ini}

An example of initialization input parameters file.

{\footnotesize
\begin{verbatim}
internal 0
internal 0
internal 0
maxNumberFeatures 12
maxNumberProcesses 9
maxNumberFeatureValues 40
initNumberPopulation 100
maxNumberBestIndivids 5
ageLimit 10
mutationFactor 0.8
yearsForEvolution 400
optimFlag 1
SignificanceChoice 3
minEventsCoefficientSignal 5.0
minEventsCoefficientBackground 5.0
weightCoef 1.0
\end{verbatim}
}

\section{User defined significance function.}
\label{sec:quality}

The whole text of the template is shown below. 

{\footnotesize
\begin{verbatim}
#include <iostream>
#include <vector>

using namespace std;

double UserDefinedQuality(const double S, const double B)
{
  // simple example of re-defined S1=S/sqrt(B)
  // using total number of weighted signal (S)
  // and sum of background (B) events

  return S/sqrt(B+0.00001);
}




double UserDefinedQuality(const double S,
                          const double B,
                          const double dS,
                          const double dB,
                          const vector<double> expEvents,
                          const vector<double> dexpEvents,
                          const vector<int> genEvents,
                          const vector<double> weights
                          )
{
  // Detailed User's Qualty function
  // available variables are:
  // S - total number of weighted signal events
  // dS - MC stat. error on total number of weighted signal events
  // B - total number of sum of weighted background events
  // dB - MC stat. error on total number of sum of weighted background events
  // expEvents - vector with weighted events (signal is the last element)
  // dexpEvents - vector with MC stat. error on weighted events (signal is the last element)
  // genEvents - corresponding numbers of generated events
  // weights - vector of weights per generated event

  // access example is demonstrated below
  if (0)
    {
      cout << "\n\nSignal events: " << S << " +- " << dS << endl;
      cout << "Background events: " << B << " +- " << dB << endl;
      for (int i=0; i<int(expEvents.size())-1; i++)
        {
          cout << "Background Process " << i
               << "\nWeighted background events " << expEvents[i]
               << " +- " << dexpEvents[i]
               << " corresponding to " << genEvents[i] << " gen. events"
               << " with weight " << weights[i]
               << endl;
        }
      int sig_index = expEvents.size()-1;
      cout << "Signal Process:"
           << "\nWeighted events " << expEvents[sig_index]
           << " +- " << dexpEvents[sig_index]
           << " corresponding to " << genEvents[sig_index] << " gen. events"
           << " with weight " << weights[sig_index]
           << endl;
    }


  // simple example
  return S/sqrt(expEvents[0]+0.0000001);
}

\end{verbatim}
}

\section{Verification data format.}
\label{sec:verif}

An example of verification data format with four different cut sets to
try (to verify).

{\footnotesize
\begin{verbatim}
4
165  3  22.2 65.5 -2.48 -2.45 0    -1 0    0    3  2
1470 11 2120 1160  2.48  2.45 4.78  1 3.14 3.14 10 8
224   3 62.8 65.5 -2.49 -2.45 0  -0.821 0   0.476 3  2
1470 11  577 1160  2.48  2.47 4.78  1 3.14 3.14 10 8
224   3 62.8 65.5 -2.49 -2.45 0  -0.821 0.494 0 3 2
1470 11 510  1160  2.49 2.47  2.98 1 3.14 3.07 10 8
150   2 22.2 65.5 -2.48 -2.45 0   -1 0    0.476 2 1
1470 11 2120 378   2.48 2.47  4.78 1 3.14 3.14 10 8
\end{verbatim}
}

\section{Penalty factors.}
\label{sec:penalty}

An example of a ``penalty'' parameters input file. In this example the
last sample, signal, gets -0.1 penalty, which means 10\% reduction in
the number of signal events.

{\footnotesize
\begin{verbatim}
0.0 0.0 0.0 0.0 0.0 0.0 0.0 0.0 -0.1
\end{verbatim}
}

\section{Switching variables ON/OFF file format, data
  errors file format.}
\label{sec:onoff}

An example of switching variables ON/OFF file format. In this example
3rd and 11th variables are switched off from the analysis.

{\footnotesize
\begin{verbatim}
1 1 0 1 1 1 1 1 1 1 0 1
\end{verbatim}
}

\end{document}